# Bridging the Gap Between Stars and Planets:

# The Formation and Early Evolution of Brown Dwarfs

A White Paper for the Astro2010 Decadal Survey

*by*

Subhanjoy Mohanty *(Imperial College London, UK)*[*]
Adam Burgasser *(Massachusetts Institute of Technology, USA)*
Gilles Chabrier *(École Normale Supérieure de Lyon, France)*
Paolo Padoan *(University of California at San Diego, USA)*
Patrick Hennebelle *(École Normale Supérieure de Paris, France)*
Ilaria Pascucci *(Johns Hopkins University, USA)*
Adam Kraus *(California Institute of Technology, USA)*
Isabelle Baraffe *(École Normale Supérieure de Lyon, France)*
Keivan Stassun *(Vanderbilt University, USA)*
Jane Greaves *(University of St. Andrews, UK)*
Ansgar Reiners *(Universität Göttingen, Germany)*
Mike Dunham *(University of Texas at Austin, USA)*
Aleks Scholz *(University of St. Andrews, UK)*
Ben Oppenheimer *(American Museum of Natural History, USA)*
Tom Ray *(Dublin Institute for Advanced Studies, Ireland)*
Daniel Apai *(Space Telescope Science Institute, USA)*
Alyssa Goodman *(Harvard University, USA)*
Kelle Cruz *(California Institute of Technology, USA)*
Louisa Rebull *(California Institute of Technology, USA)*
Estelle Moraux *(Laboratoire d'Astrophysique de Grenoble, France)*

[*] *e-mail:* s.mohanty@imperial.ac.uk. *ph.:* +44-20-7594-7553. *fax:* +44-7594-7541.

## Introduction

Straddling the divide between low mass stars and giant planets, brown dwarfs are peculiar beasts. By definition, these are *sub*-stellar bodies: with a mass < 75 $M_{JUP}$, they cannot sustain stable hydrogen fusion. Beginning life as fully convective objects, similar in temperature to M dwarf stars, they simply become (after a brief initial period of deuterium fusion) ever cooler and fainter with time, like planets, with interiors eventually supported by electron degeneracy pressure. Predicted to exist nearly half a century ago [1, 2], the first brown dwarfs were detected only in 1995 [3, 4]. Their critical relevance to stellar and planetary physics was immediately grasped however, and the Astro2000 Decadal Survey identified the discovery and characterization of large numbers of brown dwarfs as one of the decade's major goals. Wide-field red and near-infrared imaging surveys such as 2MASS, DENIS, SDSS, UKIDSS and CFHTLS have now made this a spectacular reality, revealing hundreds of these diminutive bodies in the solar neighborhood as well as in nearby star-forming regions and young clusters. Indeed, far from being mere curiosities in the galactic menagerie, brown dwarfs turn out to be some of our most ubiquitous neighbors, comparable in number density to low-mass stars [5, 6, 7]. This cornucopia of cool objects has already necessitated the invention of two new spectral classes (L and T; [8]) – the first since Morgan and Keenan laid out their spectral sequence over fifty years ago [9] – and propelled brown dwarf research in directions, and to levels of complexity, entirely unanticipated at the turn of the millennium.

In this White Paper, we focus in particular on 2 central themes that have emerged in the study of *young* brown dwarfs, and that we foresee being major drivers of stellar, brown dwarf and exo-planet research over the next decade.

1) The formation of brown dwarfs poses a fundamental mystery, key to unraveling the general physics underlying low-mass star formation and the stellar Initial Mass Function.

2) Disks girdling young brown dwarfs offer unique insights into general disk properties, disk evolution, and planet formation; they moreover raise the fascinating possibility of forming low-mass planets around brown dwarfs as well.

As these issues imply, the importance of research into brown dwarfs stems from their *direct relevance to fundamental questions of stellar and planetary origins and properties*. We address these themes below to identify the prime attendant science goals, both observational and theoretical, for the next decade.

### 1. What is the Formation Mechanism of Brown Dwarfs?

Since the seminal work of Salpeter [10], much labor has been devoted to deciphering the physical mechanisms underlying the stellar Initial Mass Function (IMF). This is a cornerstone of astrophysics, since the IMF provides the key link between stellar and galactic evolution, and determines the baryonic content and chemical enrichment of galaxies. However, a satisfactory general theory of the IMF remains elusive, and one of the major open questions in astrophysics. The unique difficulties posed by BD formation make the issue even more challenging, sharply restricting the arena of potentially viable IMF theories. The task for the coming decade is to both advance these theories to reflect nature more faithfully, as well as to observationally discriminate between them.

*Theory*

The central dilemma of BD origins is easily framed: *The very low Jeans mass required to form BDs implies gas densities far in excess of the mean in molecular clouds*. Two competing solutions to this conundrum have emerged. In the 'ejection' scenario, the requisite high densities are achieved in the interior of cores during opacity-limited collapse, forming sub-stellar mass 'stellar embryos'; BDs are those embryos ejected from the core by N-body interactions before accreting enough gas to grow into full-fledged stars [11]. The theory also predicts a relative paucity of BD binaries, especially wide ones due to truncation during ejection. SPH simulations qualitatively support this picture [12, 13]. Most, however, use an unrealistic barotropic equation of state: these vastly overestimate the relative number of BDs, while yielding BD binary frequencies of ~20% [14]. Conversely, simulations with radiative feedback and magnetic fields produce far fewer BDs, but are not large enough for statistically meaningful comparisons to the IMF; magnetic fields also severely inhibit binary formation [15, 16, 17, 18]. Intensive SPH simulations, incorporating both radiative feedback and magnetic fields and large enough to yield a statistically significant IMF, are a prime goal for rigorously testing the theory.

In the second, 'turbulent fragmentation' scenario, the high densities arise as fluctuations induced by supersonic turbulence in molecular clouds. The mass spectrum of gravitationally bound cores is set by the spectrum of turbulent velocities; stars form directly out of the larger bound cores and BDs out of the smallest, sub-stellar mass ones. The analytic theory predicts an overall IMF consistent with data [19, 20]. A robust prediction of the BD IMF, though, still faces two major hurdles: the extreme sensitivity of the turbulent fragmentation to the gas thermodynamics [21] and to the forcing of the turbulence [22]. The first requires that numerical simulations properly model cooling and radiative processes and magnetic fields; the second demands accurate modeling of the mechanisms sustaining turbulence in clouds. Also, the theory as yet makes no quantitative binarity predictions; research into this area is essential for comparison to observations. Finally, numerically modeling this theory remains a challenge, given the very small spatial scales of the density fluctuations that produce BDs. The next generation of large-scale AMR simulations, incorporating recent developments coupling radiation transfer and MHD, hold out the best promise for achieving this goal.

*Observations*

**Bound Substellar Cores:** Clearly, the existence/absence of gravitationally bound sub-stellar mass cores would strongly support/belie the role of turbulent fragmentation in shaping the IMF. While a number of sub-stellar cores have now been detected, however, their bound nature is not yet established: the majority have been identified only through sub-mm dust continuum emission [e.g. 23], without line-width data to test their virality.

Only ~10 such cores have line data, and half appear bound [24, 25; Fig.1], but the large uncertainties in both the enclosed and virial masses for any individual core (factors of up to ten, stemming from unknowns in e.g., abundances, density profiles and temperature), combined with the very small number statistics, make these results highly tentative. What is essential now is the identification of a *large* sample of sub-stellar cores, followed by multi-wavelength, resolved spectroscopy to firmly establish their bound or transient condition. Both goals should become feasible soon. The advent of SCUBA-2 on JCMT, and the planned Cornell Caltech Atacama Telescope (CCAT), will enable rapid and sensitive surveys for cores down to very low sub-stellar masses via sub-mm dust continuum imaging. CCAT's anticipated multi-wavelength capability can further provide well-constrained dust temperatures. Spectroscopic follow-up on ALMA in a suite of lines such as $N_2H^+$, $HCO^+$ and CO will yield gas mass, velocity, abundance and depletion information; ALMA can also spatially *resolve* the density structure of even very small sub-stellar cores. In combination, these data can rigorously test the existence of bound sub-stellar cores, vastly improving our understanding of the physics governing the IMF.

**Binarity:** Binary frequency and properties offer another test of BD formation and IMF theories. Currently, resolved imaging and radial velocity (RV) monitoring indicate very rough BD binary fractions of $\varepsilon_b$ ~ 20-25% in the field and ~20% in young star-forming regions (SFRs) [26–34]. Also, while the field BD binaries evince a robust peak at a mass ratio $q$ ($\equiv M_2/M_1$) ~ 1, and a sharp frequency cutoff for $\Delta > 15AU$, the young systems, all in low-density regions, have a *statistically significant* flatter $q$ distribution and a much larger spread in separations, up to ~800 AU [35, 34]. Finally, the data indicate a smooth trend of increasing $q$, decreasing separation and decreasing frequency with declining primary mass, all the way from solar-mass stars to BDs [32, 30]. These results offer some tantalizing clues. First, while the overall BD binary fractions agree with the (barotropic) SPH predictions of 'ejection', the very wide young binaries argue against the theory. Second, the disparities in $q$ and separation between the field and young systems may arise *if* most field stars/BDs are born in dense clusters (e.g., [36]) *and* post-formation dynamical evolution plays a greater role in such regions than in the low-density ones where the known low-$q$/wide systems reside. Third, the smooth trend in binary properties hints at a universal binary formation mechanism from stars to BDs.

These tentative implications, while critical for understanding BD, IMF and binary origins if borne out, are however presently based on exceedingly poor statistics. Only 9 young BD binaries are known, from imaging and intensive-RV studies of just ~35 and ~10 young sources respectively; the field surveys are similarly patchy. Statistical modeling of these sparse data implies that the true BD $\varepsilon_b$ may be twice as high as currently detected (and completely inconsistent with 'ejection'), with half at $\Delta < 2.5AU$, a barely studied regime accessible only to RV observations [37]. Constraining BD binary statistics, especially in young regions, is thus a major goal for the next decade. This will necessitate optical/NIR high-resolution spectroscopic RV surveys over >3 epochs, on 8-10m class telescopes, of a significant fraction of the ~100 young BDs known in low *and* high density SFRs and nearby young associations, as well as laser-guide-star AO imaging of the same sources. The Thirty Meter Telescope (TMT) will make this particularly feasible by greatly enhancing the achievable sensitivities, spatial resolution and sample sizes.

Furthermore, low-*q* binary formation presents a novel way of creating *planetary-mass* objects (i.e., BDs below the D-burning limit: $M < 12\ M_{JUP}$) around BD primaries. Indeed, one young 25 $M_{JUP}$ BD *does* have a 5-8$M_{JUP}$ resolved planetary-mass binary companion (2MASS 1207B) [28, 38, 39, 40], and two of the other known young low-*q* BD systems also have companions close to the D-burning boundary [33, 34]. Such young planetary-mass companions, if resolved, can provide sorely needed constraints on our theories of the initial atmospheric features and early evolution of giant planets, as has been the case with 2MASS 1207B [40]. This provides additional impetus to AO imaging surveys for wide low-*q* young BD binaries over the next decade.

Finally, in the interplay between binarity and IMF in the BD regime, *eclipsing binaries* (EBs) occupy a special position. First, they probe very tight systems. Second, they directly yield precise component masses and thus a model-independent *q*. Third, by supplying empirical masses, radii and the ratio of component temperatures, they allow a calibration of the *observed* IMF of young BDs, which can be uncertain by up to 50% due to uncertainties in the theoretical evolutionary tracks used to infer masses for very young isolated BDs [41-46]. The first and so far only known BD EB, discovered recently in the 1-Myr-old ONC region [31], illustrates this beautifully, by revealing a wholly unexpected $T_{eff}$ reversal between the components, probably due to magnetic field effects and with critical implications for the masses currently derived for very active BDs [47-50]. Over the coming decade, the sample of BD EBs will increase rapidly through ongoing large-scale variability surveys, including exoplanet transit surveys which by virtue of their design will specifically provide bright objects amenable to spectroscopic follow-up. A comprehensive investigation of these via optical/NIR high-resolution spectroscopy on 8-10m class telescopes will provide constraints on both binarity and the true BD IMF.

## 2. What do Brown Dwarf Disks imply for general Planet Formation?

It is now firmly established that BDs are commonly girdled by accretion disks in their youth: the evidence is both indirect (UV to IR spectroscopic signatures of disk-accretion) and direct (NIR to mm thermal dust emission). The fraction of BDs harboring primordial disks is comparable to that in stars [51]; moreover, the spectral energy distribution of these BD disks can be successfully modeled by scaled-down versions of stellar T Tauri disks [e.g., 52, 53, 54, 55]. By presenting an environment conspicuously different from stellar disks (scaling has physical consequences!), sub-stellar disks offer a unique testbed for theories of disk evolution and planetary origins.

*Primordial Accretion Disks*

**Disk Masses:** The masses of primordial disks critically influence the efficiency and timescale of planet formation. Estimating protoplanetary disk masses is however difficult even for solar-type stars: most of the disk mass is cold (at large radii) and resides in gas, and the continuum emission in the appropriate sub-mm/mm regime is also faint and declines sharply as $F_\nu \propto \lambda^{-4:-2}$. Gas line detections in the mm are sparse and confined to the most massive T Tauri and Herbig Ae/Be stars [56]; in their absence, stellar disk masses are estimated from sub-mm/mm dust continuum emission assuming: optically thin dust; a fiducial dust opacity; and a canonical ISM gas-to-dust ratio of 100. These assumptions make current disk mass estimates severely uncertain for stars of all types. For faint BD disks, even detecting the dust continuum is a challenge, with fluxes of order a mere mJy or less: only 6 BD disks have firm mm detections so far [57, 58]. Large-scale surveys of SFRs with JCMT/SCUBA-2 and CCAT are thus a major goal for the coming decade: these instruments can detect sub-mm dust emission from many more BD disks than possible now, and a few strong molecular lines in the most massive of these. In parallel, ALMA can extend these measurements into the mm, detect even fainter/lower-mass BD disks, and spatially resolve the brightest of the disks [59; Fig.2]. Combining spatially resolved disk emission with the slope of the SED will be particularly important for inferring typical grain sizes – and hence dust opacities – in the outer disk and thereby resolving one of the main degeneracies in estimating disk masses [60]. Together, the data will allow far better estimates of BD disk masses, and their planet-forming capacity.

**Disk Structure and Evolution:** Circumstellar disks are constantly evolving, with their primordial mass cleared out by a combination of accretion onto the central star, photoevaporation, and planet formation [61]. The grain growth that ultimately leads to planets also dramatically modifies the disk structure, by depleting the population of sub-micron grains while producing larger aggregates that more easily decouple from the gas and settle toward the disk midplane [62]. Recent Spitzer observations indicate that all these evolutionary processes are strongly dependent on the stellar mass. First, optically thick primordial disks appear to persist longer around late-M stars and BDs than around solar-mass ones [63, 64]. Second, not only do BD disks evince cleared inner holes, grain growth and dust settling, implying that the first steps towards planet formation are occurring in them just as in stellar T Tauri disks [65, 66, 67], but there is also burgeoning evidence that grain growth and settling in fact occur faster in BD disks than in stellar ones [65, 68, 69; Fig.3]. These findings hint that *disk-dispersal mechanisms other than planet formation occur less efficiently around very low mass stars and BDs*. Indeed, young BDs are the best targets to test the limits of photoevaporation in clearing inner holes, since their chromospheric UV ionizing flux is $\sim 10^3$ times less than in T Tauri stars, but their disk and central masses are each an order of magnitude smaller as well [70, 67]. Progress on this crucial issue requires the identification of a *large* sample of BD disks with inner regions cleared of dust; detailed radiative transfer modeling to infer the size of the dust cavities; and sensitive measurements of the ionizing UV flux from accreting and non-accreting young BDs. JWST will be vital for the identification of dust inner holes, while COS on HST will be invaluable for measuring the FUV ionizing continuum.

A firm estimate of the UV field is also essential for characterizing the disk chemistry and the formation and survival of complex organics in it, with key implications for the bulk composition of planets and the delivery of organics to them post-formation. Simple organic molecules have now been found in both T Tauri and BD disks [71, 72, 69], and already imply that organics in planet-forming regions of BD disks differ from those around sun-like stars [69]. Herschel, ALMA, JWST and TMT will provide the wavelength coverage and spatial/spectral resolution to study the chemistry of many BD disks with S/N comparable to that possible now only for the most massive T Tauri ones. Finally, BD disks are also excellent laboratories for understanding the formation of high-temperature products like crystalline silicates in disks. The identification of prominent crystalline silicate peaks around 10μm in a few BD disks [65, 73, 74] already shows that their formation process must be efficient around very cool stars/BDs as well as around hotter solar-types. By detecting and modeling crystalline peaks at multiple wavelengths in a large sample of BD disks, and comparing to stellar disk results, we can investigate the crystalline mass fraction as a function of stellar mass and disk radial distance, and address whether localized crystallization processes such as shocks are at work in disks.

*Debris Disks*

The fact that primordial accretion disks are ubiquitous around BDs, and exhibit grain evolution analogous to stellar disks, raises the intriguing possibility that planets, may form around sub-stellar bodies too. Proof of such evolution would be the discovery of debris disks around BDs – second generation disks formed and constantly replenished by ongoing collisions between planetesimals, and as such a clear signature of planetary construction having proceeded to at least the stage of kilometer-sized bodies. Many such disks have been identified around AFGK stars, and an increasing number are now being found around low-mass M dwarfs as well, over mid-IR to mm wavelengths [75, 76, 77]. Unexpectedly, warm debris appears more prevalent around M stars than solar-type ones [77]. A possible reason is that lower-mass objects have less massive primordial disks with longer grain-growth timescales, and thus cannot form giant planets but efficiently produce lower-mass planets and copious planetesimals. If so, debris may be even more common in BDs. Either way, such disks around BDs would illuminate many aspects of planetesimal formation, as well as open a new frontier in planetary/habitability studies.

Debris disks are generally found around stars of age $10 - \text{few} \times 10$ Myr. For a fiducial 10-Myr-old, $50 M_{JUP}$, $0.01 L_\odot$ BD, the expected 24μm debris flux (scaling from $0.2 M_\odot$ mid-M stars in NGC 2547 with similar luminosity and Spitzer-detected debris at 24μm [77], assuming disk mass scales with central mass) is ~0.1 mJy at 150pc, the distance to the nearest SFRs. Scaling similarly from the sub-mm debris emission in the $0.5 M_\odot$ early-M star AU Mic [75], the BD would emit only ~5-50 μJy over 850-450μm at 150pc, but ~0.1-1 mJy at 30pc, the mean distance to nearby young associations and moving groups. The latter fluxes are within reach of ALMA. To date, sufficiently sensitive Spitzer 24μm surveys of few-Myr-old SFRs have turned up many primordial BD disks but no debris [64, 67], possibly because these regions are too young for such evolution. Recently however, a host of BD candidate members of nearby groups within 50pc and ages 10-30 Myr have come to light [78]; targeting these (and additional ones likely to be found via ongoing and upcoming surveys, e.g., UKIDSS and WISE) for debris disk searches is a prime goal for the coming decade. In the sub-mm/mm, the ideal instrument, with both the requisite sensitivity and high-resolution to beat extragalactic confusion, will be ALMA (confusion limit < 1μJy). In the MIR, SPICA and JWST, with sensitivities of few-10 μJy at ~20μm, can potentially detect a plethora of warm debris disks around BDs.

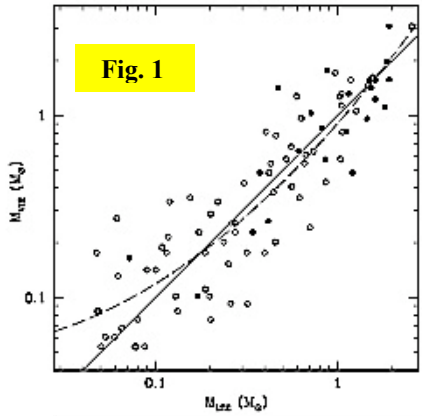

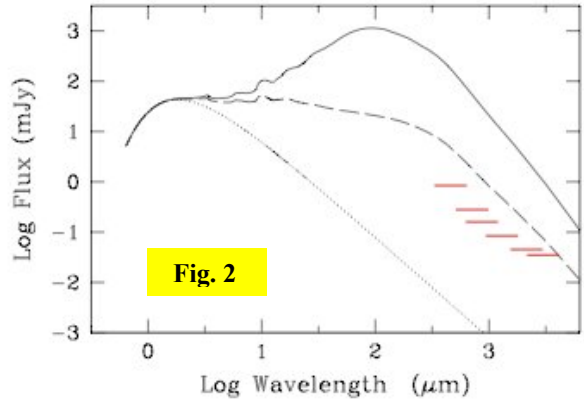

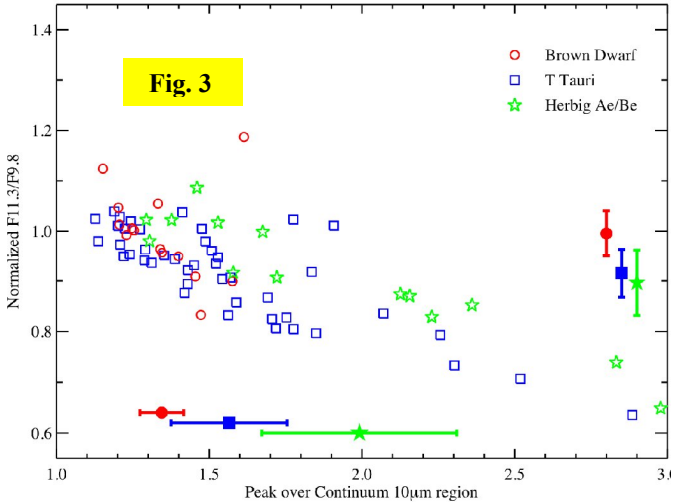

**Fig.1:** LTE mass (x-axis) versus virial mass (y-axis) for detected cores in NGC 1333, in units of solar mass. Filled circles are cores with known protostars, unfilled ones are those without. Solid line shows locus along which the two masses are equal; bound cores should lie below this line (LTE mass > virial mass). The curved line is a quadratic fit to the data. Only 10 cores with LTE masses in the BD regime are detected; 5 appear bound or close to it. From Walsh et al. 2007.

**Fig.2:** Model for young 50 $M_{JUP}$ BD at 140pc with 5$M_{JUP}$ disk (flared disk: solid, flat disk: dashed, photosphere: dotted), against ALMA sensitivities (red lines). ALMA should easily see even the flat disk at all wavelengths. From Natta & Testi 2008.

**Fig. 3:** Ratio of normalized flux at 11.3um to 9.8um (proxy for degree of disk grain crystallinity; y-axis) vs peak over continuum in 10um region (proxy for grain growth), for Herbig Ae/Be stars (green), T Tauri stars (blue) and BDs (red). Filled symbols with error bars show mean value and 1sigma deviation for each group. BD disks are the most evolved: highest crystallinity and largest grains (and thus weakest 10um emission). From Pascucci et al. 2009.